# Longitudinal quantum plasmons in copper, gold, and silver

M. Moaied,[1,2,3]*, S. Palomba[1], and K. Ostrikov[2,3,4]

[1]*Institute of Photonics and Optical Science (IPOS), School of Physics, The University of Sydney, NSW 2006, Australia*

[2]*Complex Systems, School of Physics, The University of Sydney, NSW 2006, Australia*

[3]*Commonwealth Scientific and Industrial Research Organisation (CSIRO), Lindfield, NSW 2070, Australia*

[4]*School of Chemistry, Physics, and Mechanical Engineering, Queensland University of Technology, Brisbane, Queensland 4000, Australia*

*email: m.moaied@physics.usyd.edu.au

## Abstract

The propagation of plasmonic waves in various metallic quantum nanostructures have considered attention for their applications in technology. The quantum plasmonic properties of metallic nanostructures in the quantum size regime have been difficult to describe by an appropriate model. Here the nonlocal quantum plasmons are investigated in the most important metals of copper, gold, and silver. Dispersion properties of these metals and propagation of longitudinal quantum plasmons in the high photon energy regime are studied by a new model of nonlocal quantum dielectric permittivity. The epsilon-near-zero properties are investigated and the spectrum and the damping rate of the longitudinal quantum plasmons are obtained in these metals. The quantum plasmon's wave function is shown for both classical and quantum limits. It is shown that silver is the most appropriate for quantum metallic structures in the development of next-generation of quantum optical and sensing technologies, due to low intrinsic loss.

**Introduction**

Plasmons are collective and coherent plasma oscillations of free electrons in metallic materials which play a fundamental role in the optical properties of metals [1]. This has stimulated a wide range of theoretical investigations and applications, such as optical metamaterials [2,3], solar cells [4-7], biochemical sensing [8], and antennas [9,10].

The classical Drude model is the most accepted theoretical framework to model the dielectric permittivity in metallic structures [11-13]. However, a discrepancy between the classical Drude model and the measured permittivity of quantum sized metallic nanostructures has been reported in the literature [14]. For example, it has been investigated [15] that the plasmon resonance energy in the metallic nanoparticles has a substantial deviation from classical predictions and shifts to a higher energy when the size of the structure reduces to a quantum size (1-10 nm). This is due to the quantum nature of the free charge carriers and the dynamic response of metallic nanostructures to the self-consistent electromagnetic fields [16]. Therefore, the quantum effects are important and quantum mechanics is necessary to model the behavior of charge carriers in metallic objects.

The field of quantum plasmonics has recently received considerable attention not only for fundamental science [17] but also for its potential applications in plasmonic systems (e.g., metallic nanowires [18], quantum wells [19], surface plasmon polaritons waveguides [20,21], and metallic nanoparticles [22,23]). In these systems, quantum effects such as nonlocality [23-25] are important because they are responsible for plasmon-frequency shifts, which are not explained by classical theory.

We recently have theoretically investigated [26-28] the quantum properties of plasmons in metals by applying the Wigner equation to the kinetic theory of electrons where the dominant electron scattering mechanism is the electron-lattice interaction and the nonlocal longitudinal dielectric permittivity of metals (wavenumber-dependence of metal responses) was introduced. In this letter, we focus on applying this quantum theory to the most important metals for plasmonic applications (gold, silver and copper), predicting their optical behaviour in the quantum regime. We investigate the quantum properties of the longitudinal plasmonic waves to obtain their spectrum and damping in bulk materials.

**Theory**

Investigating the plasmonic properties of metallic nanostructures requires solving the kinetic and Maxwell's equations to describe the excitation of charged particles by light. Also, with considering the density of electronic states in metals at room temperature, the quantum effects cannot be ignored [16]. Quantum effects arise from the quantum nature of free charge carriers and the dynamic response of these structures to self-consistent electromagnetic fields. Hence, it is necessary to implement a quantum mechanical treatment to model the plasmons in metallic structures.

We use the kinetic Wigner equation including the collision term [26], for the distribution function of free electrons in metals in an ion background, where collision of metal electrons occur due to lattice vibrations (electron-phonons interaction) and dominates other collisions (e.g., electron-electron). Associating the Maxwell's equations and following the standard procedure described in Ref. 26, the nonlocal longitudinal dielectric permittivity of metals has been introduced as $\varepsilon(\omega,k) = \varepsilon_\infty + \chi(\omega,k)$ where

$$\chi(\omega,k) = \frac{\dfrac{3\omega_p^2}{4k^2 V_F^2} \left\{ \begin{array}{l} 2 - \dfrac{m}{\hbar k^3 V_F}\left[ k^2 V_F^2 - \left(\omega+i\nu+\dfrac{\hbar k^2}{2m}\right)^2 \right] \ln\left| \dfrac{\omega+i\nu-kV_F+\dfrac{\hbar k^2}{2m}}{\omega+i\nu+kV_F+\dfrac{\hbar k^2}{2m}} \right| \\ + \dfrac{m}{\hbar k^3 V_F}\left[ k^2 V_F^2 - \left(\omega+i\nu-\dfrac{\hbar k^2}{2m}\right)^2 \right] \ln\left| \dfrac{\omega+i\nu-kV_F-\dfrac{\hbar k^2}{2m}}{\omega+i\nu+kV_F-\dfrac{\hbar k^2}{2m}} \right| \end{array} \right\}}{\left[ 1 - \dfrac{i\nu}{2kV_F} \ln\left| \dfrac{\omega+i\nu+kV_F}{\omega+i\nu-kV_F} \right| \right]}, \quad (1)$$

is the susceptibility of electrons, $\varepsilon_\infty$ is the offset value of the permittivity, $\omega_p = \sqrt{4\pi n e^2/m}$ is the plasma frequency of the electrons, $V_F = \sqrt{2E_F/m}$ is the Fermi velocity of the electrons, $\omega$ is the frequency, $k$ is the longitudinal plasmon wave number, $\nu$ is the collision frequency describing the electron-lattice collisions, $\hbar$ is the reduced Planck constant, and $m$, $e$, and $n$ are the effective mass, charge, and number density of electron, respectively. In the classical limit when $kV_F \ll \omega$, Eq. (1) reduces to the Drude model

$$\chi(\omega) = -\frac{\omega_p^2}{\omega(\omega+i\nu)}, \quad (2)$$

in metals. Therefore, one can thus expect stronger quantum effects at higher $k$, which leads to the spatial nonlocality of electron responses in the metal. Table 1 shows the electrons parameters for copper, gold, and silver obtained from matching of the classical Drude model and experimentally measured data [13,29].

**Table. 1.** Electron parameters (see Ref. 13 and 29) for copper, gold, and silver.

|    | $\varepsilon_\infty$ | $\omega_p$ (s$^{-1}$) | $V_F$ (cm/s) | $\nu$ (s$^{-1}$) | $n$ (cm$^{-3}$) | $m$ (gr) |
|----|---|---|---|---|---|---|
| Au | 9.6 | $1.37\times10^{16}$ | $1.40\times10^8$ | $1.07\times10^{14}$ | $5.90\times10^{22}$ | $9.1\times10^{-28}$ |
| Ag | 5.9 | $1.36\times10^{16}$ | $1.39\times10^8$ | $0.30\times10^{14}$ | $5.85\times10^{22}$ | $9.1\times10^{-28}$ |
| Cu | 10 | $1.34\times10^{16}$ | $1.04\times10^8$ | $1.50\times10^{14}$ | $8.46\times10^{22}$ | $13.6\times10^{-28}$ |

Figure 1 shows the nonlocal longitudinal dielectric permittivity of copper, gold, and silver in a broad range of energies and wavenumbers. The dashed lines correspond to the epsilon-near-zero condition, $\text{Re}[\varepsilon(\omega,k)] = 0$, when the resonance frequency approaches to the bulk plasmon resonance of noble metals. These dashed lines in Fig. 1 determine longitudinal quantum plasmon resonances that have two different modes for the given $k$. The low-frequency mode starts from the infrared and the terahertz ranges and increases monotonically with $k$. This mode of resonance cannot propagate, due to the high absorption in metals.

On the other hand, the high-frequency mode, which is excited in a weak damping area, propagates in the classical regime, $k \ll 1\,\text{nm}^{-1}$, with the classical energies $2.79$, $2.91$, and $3.65\,\text{eV}$ for copper, gold, and silver, respectively. The energy of this mode increases due to quantum effects when $k$ becomes larger than $1\,\text{nm}^{-1}$ and nonlocality effects become significant. One can see a higher intrinsic loss in the nonlocal regime of the high-frequency mode. These two modes are merged together at the critical energies of $4.45$, $4.69$, and $5.73\,\text{eV}$ for copper, gold, and silver, respectively. Figure 1 also shows a different behavior of the dielectric permittivity of silver compared to copper and gold. Longitudinal quantum plasmons in silver resonate with the energy much higher and the absorption is much weaker compared to plasmon resonances in copper and gold.

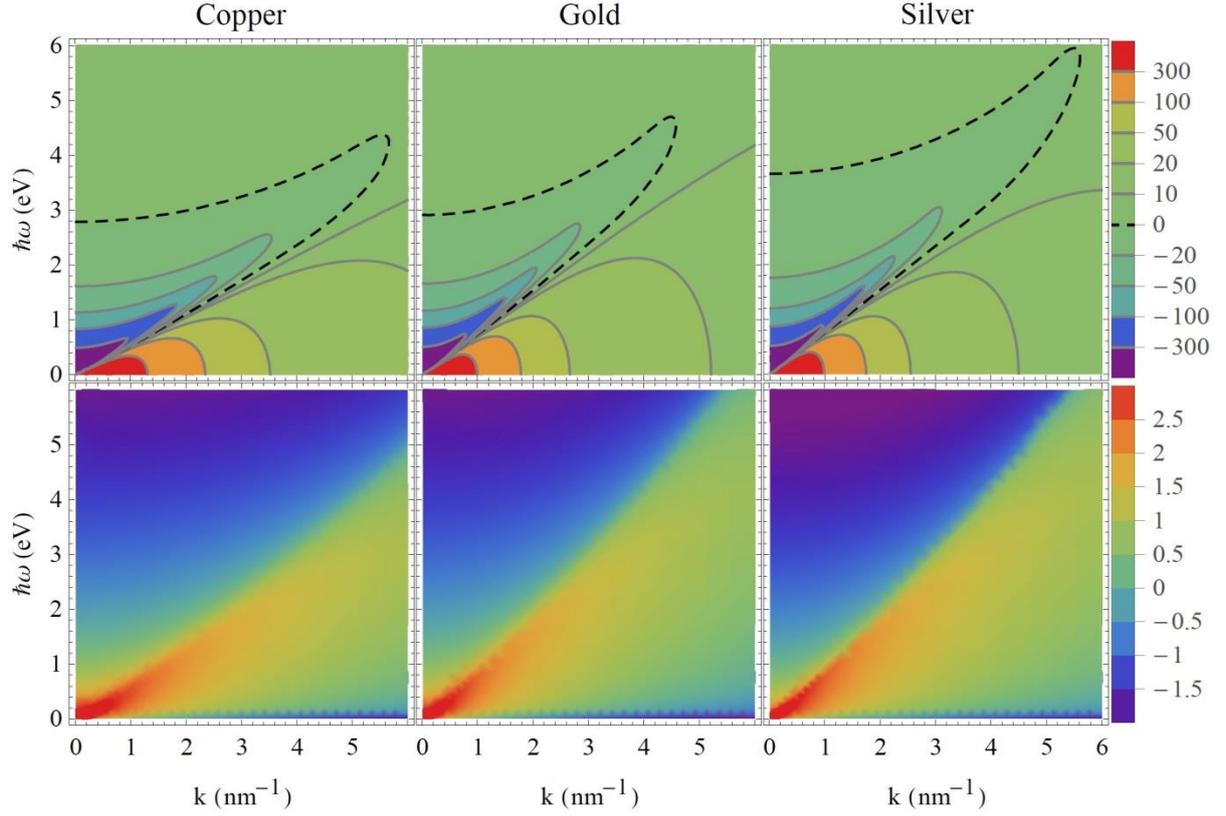

**Fig.1.** The frequency- and $k$-dependent nonlocal dielectric permittivity of metals; top $\text{Re}[\varepsilon(\omega,k)]$ and bottom $\log\left[\text{Im}[\varepsilon(\omega,k)]\right]$. Dashed line correspond to epsilon-near-zero properties of metals $\text{Re}[\varepsilon(\omega,k)]=0$.

**Dispersion of longitudinal quantum plasmons in bulk metals**

The dispersion equation, $\varepsilon(\omega,k)=0$, for bulk materials gives the spectra, $\omega(k)$, the damping rate, $\gamma(k)$, and the definition of propagation length, $L(\omega)$, of the weakly damped longitudinal quantum plasmons in copper, gold, and silver (see Fig. 2). In Fig. 2(a), the spectra of the longitudinal quantum plasmons start from the classical energies 2.79 and 2.91 eV in the classical regime and increase up to the critical energies of 4.45 and 4.69 eV in the quantum regime for copper and gold, respectively. However, the longitudinal quantum plasmons in silver are excited with the higher energies range 3.65 - 5.73 eV in silver than the above-mentioned energies ranges for copper and gold.

In the classical range, the damping rate (see Fig. 2(b)) for copper and gold is 49.8 and 35.2 meV, respectively. These values take into account the effect of the electron-lattice collisions, and increase up to 100.5 meV and 69.61 meV in the quantum regime due to the Cherenkov effect[28]. This effect is the resonant absorption of the longitudinal quantum plasmon's energy by electrons and is important when the quantum effects of the electron's motion become significant. In silver, the damping rate of the longitudinal quantum plasmons is 9.9 - 19.9 meV which is much smaller compared to copper and gold.

Figure 2(c) shows the propagation length of the longitudinal plasmons in the broad range of energy. In the classical energy range (less than 2.79, 2.91, and 3.65 eV for copper, gold, and silver, respectively), the propagation length $L$ is of the order of the atomic radius and the average inter-particle distance $n^{-1/3}$. One can see that the propagation length increases to the range of 0.5-5 and 1-10 nm for copper and gold, respectively, when the quantum effects become significant. In silver, the propagation length of the longitudinal plasmons is in the range of 3-30 nm which is much larger than those in copper and gold. This is largely due to much weaker damping rate in silver compared to other two metals.

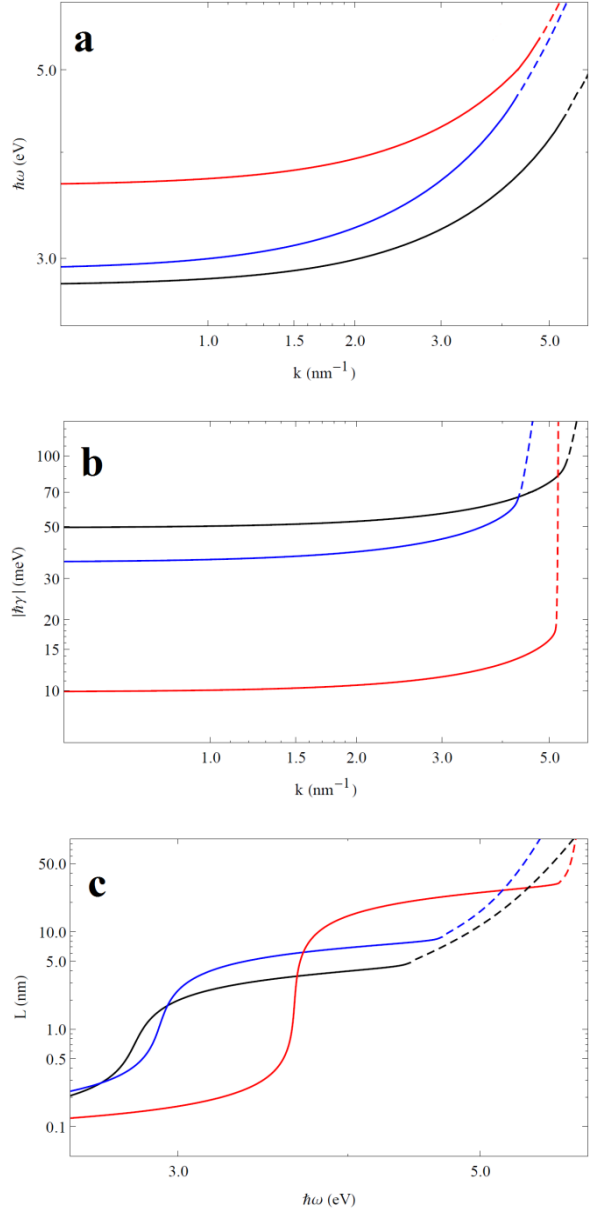

**Fig. 2.** Dispersion of longitudinal quantum plasmons in metals (orange lines: copper; yellow lines: gold; gray lines: silver). (a) spectra (b) damping rate (c) propagation length. Dashed lines correspond to strongly damped waves.

Figure 3 shows the real part of plasmon's wave function,

$$\mathrm{Re}\left[\Psi(r)\right] = A\,\mathrm{Re}\left[\exp(ikr)\right], \qquad (3)$$

in copper, gold, and silver at various energies. Here, $A$ is the amplitude of the electron's wave function at the position $r=0$. In the classical regime that the classical statistical pressure dominates the wave dynamics, the electrons can be considered as a point (black lines in Fig. 3) and the extension of the plasmon's wave function in space is smaller than or of the order of the average inter-atomic distance. This extension increases up to 20 nm beyond the average inter- atomic distance due to the quantum effects when the energy of the longitudinal quantum plasmons increases so that plasmon's wave functions overlap. Therefore, the quantum effects become important in the system and the Cherenkov absorption and the damping increases in the metals. In other words, the quantum effects become important when the Fermi energy

$$E_F = (3\pi^2 n)^{2/3} \hbar^2 / 2m, \quad (4)$$

exceeds the electrons thermal energy [30]

$$T_e = \hbar^2 / 2m\lambda_B^2. \quad (5)$$

Here $\lambda_B$ is the thermal de Broglie wavelength of electrons, which quantifies the extension of plasmon's wave function due to the quantum uncertainty. One can see the difference of the extension of the plasmon's wave function in silver in the quantum regime compared to the smaller extension in copper and gold.

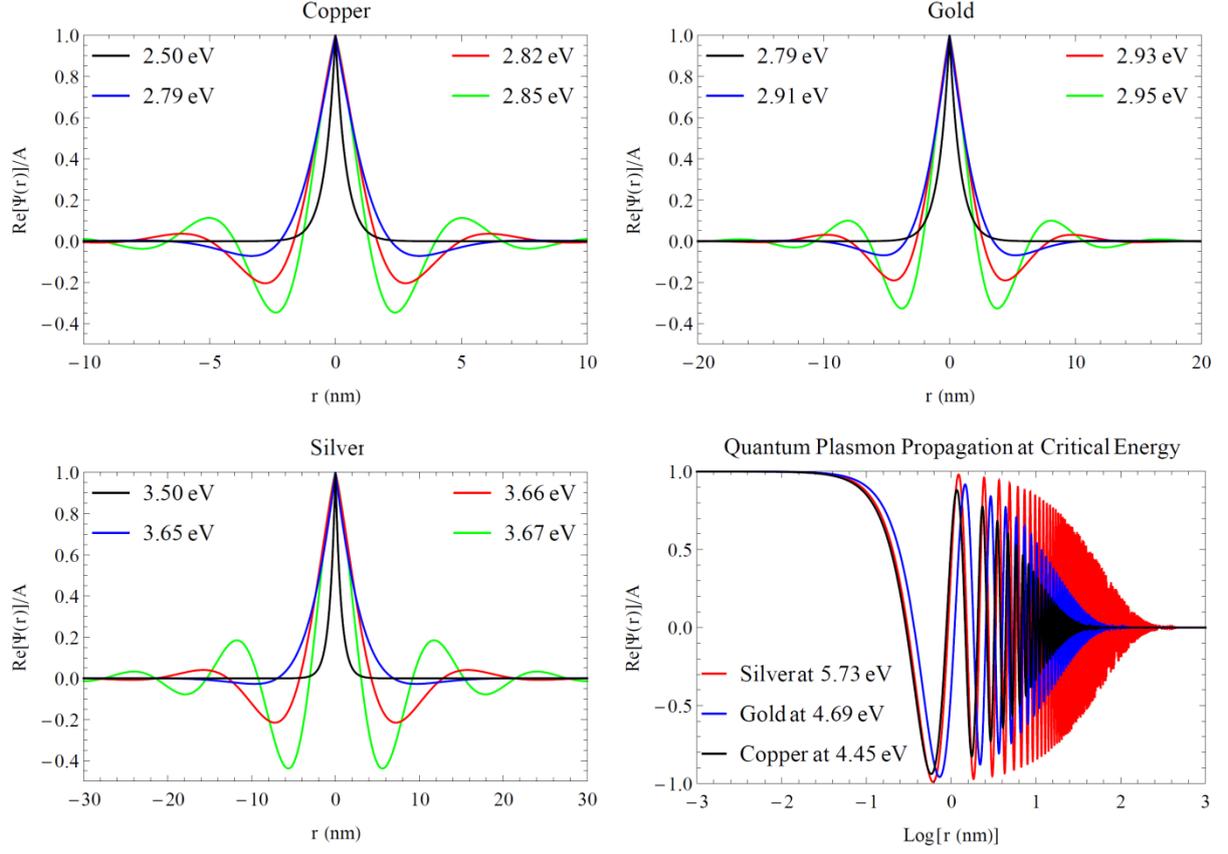

**Fig. 3.** The real part of the plasmon's wave function, $\text{Re}[\Psi(r)]$, in metals at various energies.

# Effects of non-local longitudinal quantum plasmons on propagation of electromagnetic waves in thin film slab of metals

To demonstrate validity of the theory presented in the previous sections, we perform numerical calculations of the propagation of transverse magnetic (TM) waves in the plasmonic structure using COMSOL Multiphysics. Consider the geometry [26] which shows an infinitely extended planar slab of a metal, with the permittivity defined by the quantum model of the charge carriers and the thickness $d$ which is deposited on a dielectric with a refractive index of $n_d = 1.5$ and excited by a TM plane wave in air ($n_a = 1$).

Figure 4 shows the absorption of TM waves in the metals. One can see the peak of the absorption corresponding to the high-frequency mode of the longitudinal quantum plasmons in Fig. 1. This peak does not exist at low energies (blue lines) and then appears at larger energies (see green lines) in the limit $k \ll 1\,\mathrm{nm}^{-1}$, which is corresponded to the classical excitation of plasmonic waves, and finally increases at the higher energies (red and turquoise blue lines), due to the quantum non-locality effects. In quantum regimes, the photons excite the longitudinal quantum plasmons with the wavenumbers of $k \sim 1\,\mathrm{nm}^{-1}$ or larger and the quantum non-locality peaks are much sharper than those in classical cases. The second smooth peak corresponds to the low-frequency mode of resonance. It is clear that the absorption of light by silver is much less than the absorption of light by copper and gold, due to lower damping and higher extension of the plasmon's wave function in silver. The quantum non-locality effects in silver are also more pronounced than in other metals.

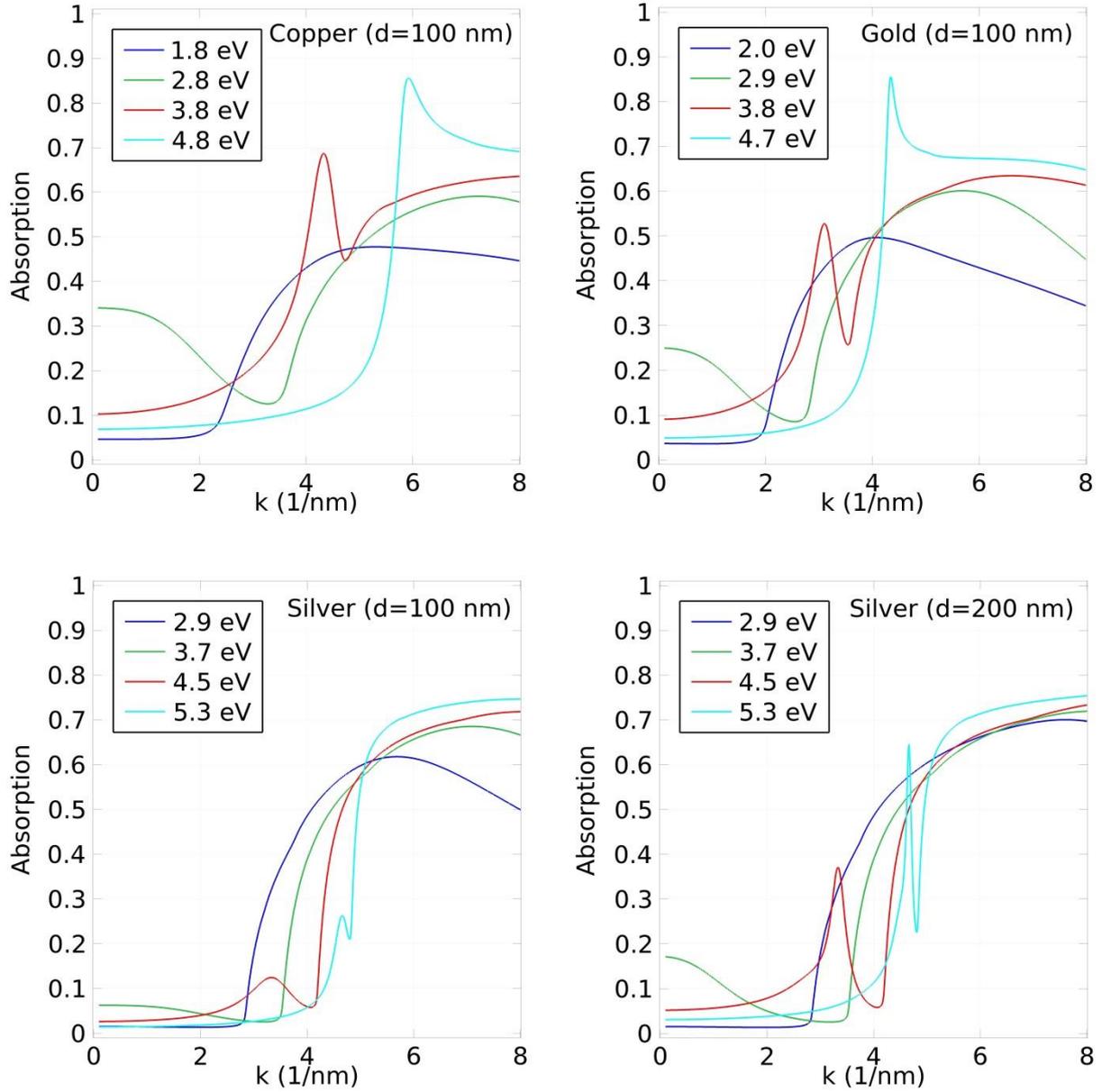

**Fig. 4.** Absorption of TM plane wave through a planar slab of metals with the thickness of $d$.

## Conclusion

In conclusion, when the energy of plasmon excitations increases above a certain threshold, quantum effects such as non-locality begin to play a crucial role to determine the propagation, damping, and losses of optical excitations in metals. Compared to the classical Drude model, which is valid at the infrared and visible wavelengths, the non-local theory is a better choice to model the permittivity of metals. These refined models provide accurate predictions for plasmon wave behaviours in the ultra-violet energy range. Our model suggest that silver, which has a low intrinsic loss, is the most appropriate for quantum metallic structures as well and for enhancing the propagation of plasmon waves at higher energy.

## Acknowledgments

M. Moaied would like to thank the University of Sydney for receiving the Australian Postgraduate Award. This work was partially supported by the Australian Research Council and CSIRO's Science Leadership Scheme.